\documentclass[twocolumn,aps,prl,showpacs,superscriptaddress]{revtex4}

\usepackage{amssymb,amsmath}
\usepackage{graphicx}
\usepackage[amssymb,pstricks]{SIunits}

\begin{document}

\title{Enhancing Optical Gradient Forces with Metamaterials}

\author{Vincent~Ginis}
\affiliation{Applied Physics Research Group (APHY), Vrije Universiteit Brussel, Pleinlaan 2, B-1050 Brussel, Belgium}
\author{Philippe~Tassin}
\affiliation{Ames Laboratory---U.S. DOE and Department of Physics and Astronomy, Iowa State University, Ames, Iowa 50011, USA}
\author{Costas~M.~Soukoulis}
\affiliation{Ames Laboratory---U.S. DOE and Department of Physics and Astronomy, Iowa State University, Ames, Iowa 50011, USA}
\author{Irina Veretennicoff}
\affiliation{Applied Physics Research Group (APHY), Vrije Universiteit Brussel, Pleinlaan 2, B-1050 Brussel, Belgium}

\date{\today}

\begin{abstract}
We demonstrate how the optical gradient force between two waveguides can be enhanced using transformation optics. A thin layer of double-negative or single-negative metamaterial can shrink the interwaveguide distance perceived by light, resulting in a more than tenfold enhancement of the optical force. This process is remarkably robust to the dissipative loss normally observed in metamaterials. Our results provide an alternative way to boost optical gradient forces in nanophotonic actuation systems and may be combined with existing resonator-based enhancement methods to produce optical forces with an unprecedented amplitude.
\end{abstract}

\maketitle

The momentum of light has been a fascinating subject for scientists for several centuries \cite{Newton-1704, Maxwell-1873} and even today the true nature of the photon's momentum raises a lot of attention \cite{Barnett-2010}. In everyday life, this momentum is normally too small to have any significant effect, but in nanoscale devices the transfer of linear momentum between light and matter and 
the associated optical forces start to play an increasingly important role \cite{Antonoyiannakis-1999}. Generally, these forces can be divided into scattering 
and gradient forces, depending on whether the transferred momentum is parallel or perpendicular to the direction of propagation. Optical scattering forces have been used to cool down atoms through the interaction with intense laser light \cite{Cohen-1998} and, more recently, for the generation of tractor beams \cite{Chan-2011} and in the field of cavity optomechanics where the coupling between the optical and mechanical modes of a cavity is exploited to manipulate the vibrations of a mechanical system \cite{Kippenberg-2008,Bagheri-2011}. Optical gradient forces are used in optical tweezers, where microscopic dielectric particles are trapped and moved by laser beams towards regions of highest intensity \cite{Ashkin-1986}.

Recently, optical forces have been studied acting upon metamaterial constituents to manipulate these elements on a mesoscopic level \cite{Tassin-2010,Kivshar-2012, Zheludev-2012,He-2012}. On the macroscopic level, it has been suggested that optical forces could be employed for all-optical device actuation, since they can generate measurable displacements in nanophotonic, optomechanical systems \cite{Povinelli-2005,Li-2008, Wiederhecker-2009,VanThourhout-2010}. Indeed, when two waveguides are closely spaced apart, the interaction of the evanescent waves of one waveguide with the other generates an optical force directed perpendicular to the waveguides. The force can be repulsive or attractive depending on the relative phase of the optical fields in the waveguides \cite{Povinelli-2005,Roels-2009}. Although the optical force---with a typical magnitude of the order of piconewtons per milliwatt---is large enough for exciting experiments in optomechanics, larger forces would be favorable for photonic systems involving optical device actuation \cite{Povinelli-2004,Eichenfield-2007,Zhang-2011}.
\begin{figure*}
  \begin{center}
    \includegraphics[clip]{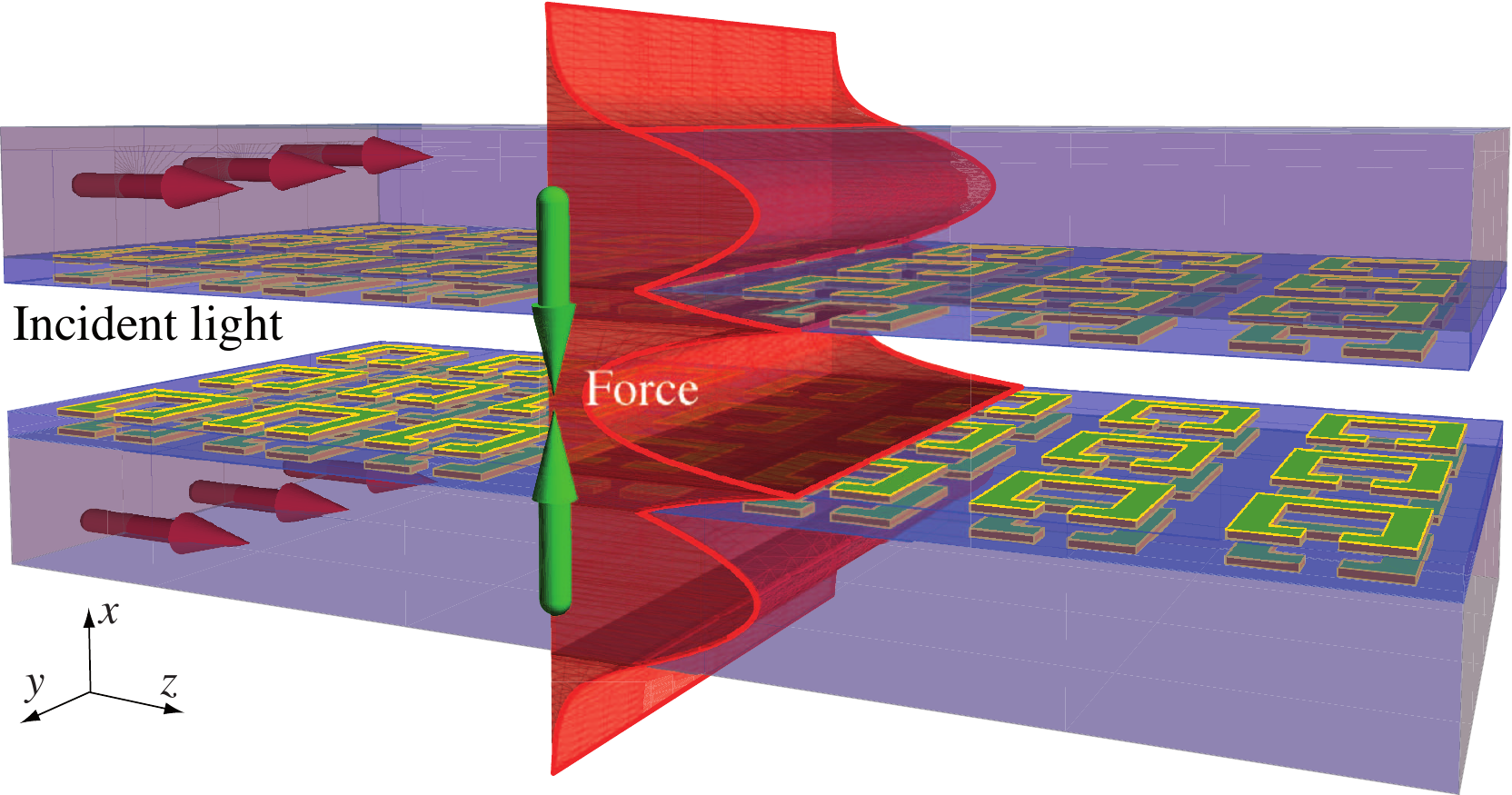}
  \end{center}
  \caption{The setup to enhance the optical gradient force between two dielectric slab waveguides. A metamaterial cladding (indicated by the split-ring resonators) is derived from a folding transformation to reduce the distance between the two dielectric waveguides (shown in plain gray) perceived by light. Since the optical gradient force decays with distance, the metamaterial slab allows enhancing the force.}
  \label{Fig1}
\end{figure*}

In this Letter, we propose a novel mechanism to enhance optical gradient forces using the method of transformation optics. The idea originates from the observation that the optical gradient force between two waveguides decays exponentially with the distance between the waveguides \cite{Povinelli-2005}. Transformation optics, however, allows us to engineer the interwaveguide distance perceived by light \cite{Pendry-2006,Leonhardt-2006} . We achieve this result by designing a medium that annihilates the optical space between two objects by implementing a folded coordinate transformation \cite{Lai-2009,Leonhardt-2009,Chen-2010} (see Supplemental Material \cite{supplement}). We study two waveguides separated by such an annihilating optical medium. For symmetry purposes and practical feasibility, the medium can be attached to the interior boundaries of both waveguides, as shown in Fig.~1. The optical forces between the two waveguides are calculated using the Maxwell stress tensor formalism. 

In Fig.~2, we evaluate the optical force acting upon one of the waveguides as a function of the separation distance. We find excellent agreement between the force acting upon the transformed waveguides and the force acting upon two traditional waveguides without metamaterial cladding---positioned at a closer distance $d_\mathrm{w/o} = d_\mathrm{with} - 2t$, where $t$ is the thickness of the annihilating coating. This relation between the physical distances $d_\mathrm{w/o}$ and $d_\mathrm{with}$ is determined by the thickness $t$ and the material parameters $\epsilon$ and $\mu$ of the metamaterial. For instance, to annihilate a distance that is $\alpha$ times larger than the thickness $t$ of the metamaterial slab, the required permittivity and permeability components in the case of TE polarization are $\mu_{xx} = -1/\alpha$, $\mu_{yy} = -\alpha$ and $\epsilon_{zz} = -\alpha$. It is important to note that the equivalence shown in Fig.~2 is valid for any transformation-optical medium that changes the gap distance between the waveguides or a transformation that changes the thickness of the waveguide's core. As such, the analogy between the two setups offers an elegant framework to determine the optical force between waveguides with complex material parameters. This also demonstrates that transformation optics remains valid if the design or calculation of optical forces is involved.

\begin{figure}
  \begin{center}
    \includegraphics[clip]{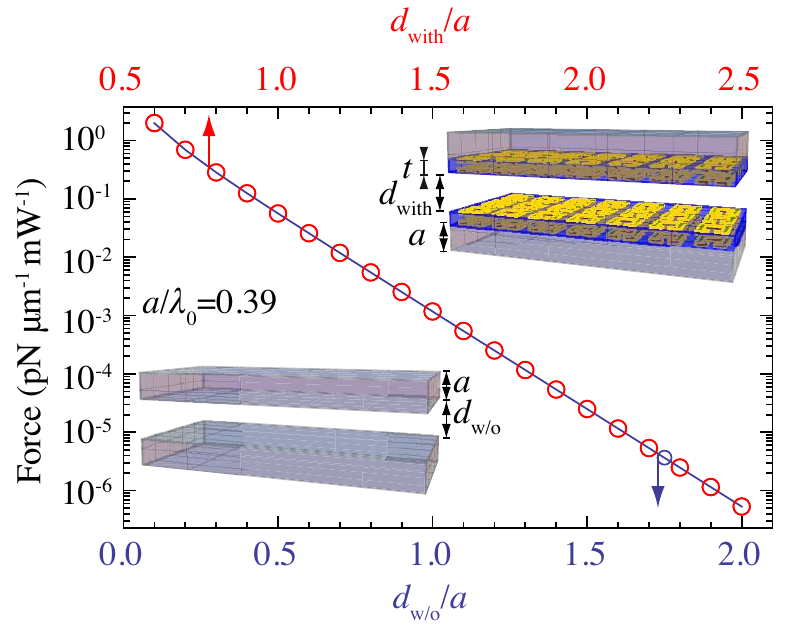}
  \end{center}
  \caption{Equivalence between the optical force in the traditional configuration of two silicon slab waveguides with $n = 3.476$ and the transformed setup of silicon waveguides ($n = 3.476$) with a metamaterial medium attached to it ($n = -1$, $t = 0.25a$). The solid blue line shows the optical force between two bare waveguides as a function of the gap distance $d_\mathrm{w/o}$. The red circles represent the optical force generated in the transformed configuration as a function of the gap between the metamaterial slabs $d_\mathrm{with}$. The metamaterial slab is designed such that $d_\mathrm{with} = d_\mathrm{w/o} + 0.5a$. In these simulations, the fundamental TE modes are considered for $a/\lambda_0 = 0.39$, corresponding to a typical setup where $\lambda_0 = \unit{1.55}{\micro\meter}$ and $a = \unit{600}{\nano\meter}$.}
  \label{Fig2}
\end{figure}

The physics of the transformation of the interwaveguide distance as shown in Fig.~2 can be further understood by looking at the field profile of the electromagnetic modes. The annihilating metamaterial in between the two waveguides acts as a perfect lens \cite{Pendry-2000} that amplifies the evanescent tails of the waveguide mode before they decay in the air gap [compare Fig.~3(a) with Fig.~3(b)]. The evanescent tails therefore interact in the same way as if they were at a closer distance, resulting in an identical electromagnetic force. A possible implementation of the left-handed metamaterial would be the structure shown in Fig.~1. The unit cell of this metamaterial consists of a double layer of split-ring resonators. The permittivity in the $z$ direction and the permeability in the $x$ direction can be generated by, respectively, the electric and the magnetic moments of the individual split-ring resonators. In addition, a magnetic moment can be generated in the $y$ direction due to the mutual interaction of the ring resonators acting as a wire pair with circulating currents in the $xz$ plane \cite{Smith-2004,Engheta-2006,Shalaev-2007,Sinclair-2010,Soukoulis-2011}.

In order to compare the magnitude of the optical force between a metamaterial-enhanced setup and a traditional setup, two important effects have to be taken into account. First, light has to be coupled into the structure, e.g., from a waveguide without the metamaterial cladding [as illustrated in Fig.~4(a)]. This will reduce the amount of power launched into the enhanced waveguides because of reflection and scattering into leaky modes. Second, the metamaterial cladding will result in extra losses that dissipate some of the input power. In Fig.~4(b), we plot the optical force on a \unit{10}{\micro\meter}-long section of the waveguide as a function of the gap distance for several implementations of the metamaterial cladding. As a reference, we also show the result for the traditional configuration without metamaterial cladding (black line), which is consistent with the results reported in the literature \cite{VanThourhout-2010}. For a fair comparison, the gap distance in the latter configuration is chosen to be the same as the distance between the metamaterial claddings in the enhanced setup (compare the insets of Fig.~2).

The blue curve (squares) in Fig.~4 corresponds to a configuration with a \unit{100}{\nano\meter}-thick left-handed metamaterial slab ($\epsilon$, $\mu = -1$) with a loss tangent of $10\%$. Two regimes can be distinguished in this curve. The linear regime (gap larger than \unit{200}{\nano\meter}) is a shifted version of the curve corresponding to the traditional configuration without metamaterials. This reconfirms the transformation-optical enhancement of the force by way of the metamaterial slabs, even when realistic losses are included. For gap distances smaller than \unit{200}{\nano\meter}, the force decreases again. Since the two left-handed slabs of \unit{100}{\nano\meter} each completely annihilate the inner gap when the gap distance equals \unit{200}{\nano\meter}, the optical force is maximized at this point. A further reduction of the gap results in an overcompensation and decrease of the optical force. This result is consistent with the fact that the largest optical force between two slab waveguide is found when they are positioned adjacent to each other and thus limits the maximum optical force amplification between two waveguides for a given input power.
\begin{figure}
  \begin{center}
    \includegraphics[clip]{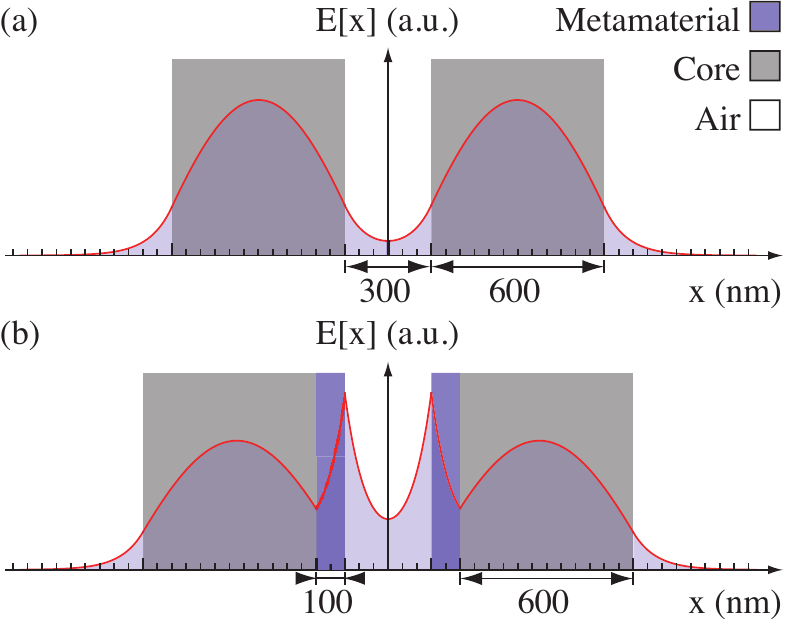}
  \end{center}
  \caption{Electric field profiles of the fundamental mode in the waveguide pair. (a)~The traditional configuration of two slab waveguides (indicated by the gray rectangles). If the waveguides are placed close to each other, the evanescent tails interact with the waveguides, resulting in a gradient force between the two slabs. (b)~The configuration with a metamaterial cladding as annihilating medium. In between the slab waveguides, we place a metamaterial slab that amplifies the evanescent tails. We demonstrate that this effect enhances the optical gradient force considerably.}
  \label{Fig3}
\end{figure}

The maximum amplification in the waveguide pair with a double-negative metamaterial as the annihilating medium is somewhat limited by the relatively high loss tangent of the double-negative metamaterial. This dissipative loss is closely related to the resonant constituents required to achieve negative permeability \cite{Tassin-2012}. To overcome this limitation, we propose to implement the annihilating slab using a poor man's lens; i.e., we replace the double-negative metamaterial with a single-negative metamaterial \cite{Pendry-2000}. This is a legitimate approximation if we work with TM-polarized light, for which only the permittivity of the metamaterial determines the mode profile and the resulting optical forces. Since the slabs no longer require negative permeability, we can use nonresonant metamaterials with smaller loss tangents. For example, for a poor man's lens implementation of the annihilating medium with $\epsilon = -1$ and loss tangent of $10\%$, we observe in Fig.~4(b) (red circles) that the maximal optical force enhancement equals $16.4$, which is actually larger than with the double-negative metamaterial cladding. When we reduce the loss tangent to $5\%$ (red triangles) and $1\%$ (red diamonds), as appropriate for single-negative materials, we reach enhancement factors of up to $243$. The use of single-negative metamaterials, which can be implemented with a stack of thin metal sheets, has the added advantage of low dissipative loss, so that thermomechanical forces will be unimportant. Stacked metal sheets have a similar area moment of inertia (compare the Young's moduli for silica and gold), i.e., their mechanical properties will not be adversely affected by the metamaterial structure.
\begin{figure}
  \begin{center}
    \includegraphics[clip]{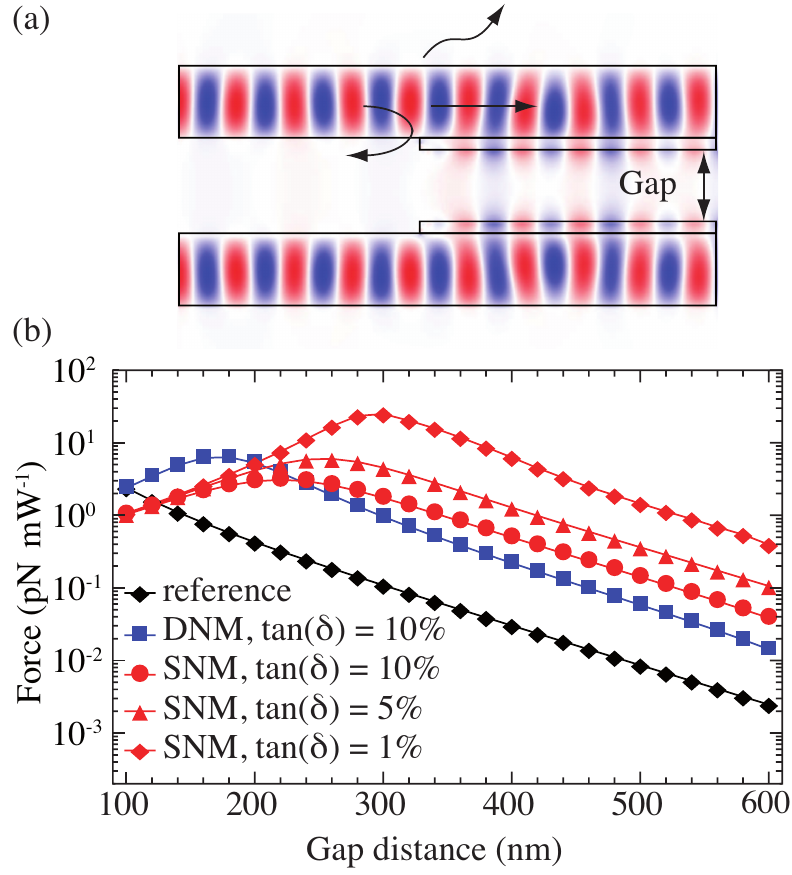}
  \end{center}
  \caption{The optical force on a \unit{10}{\micro\meter}-long waveguide section as a function of the gap distance between the waveguides for TM-polarized light ($\lambda_0 =  \unit{1.55}{\micro\meter}$, $P_\mathrm{input} = \unit{1}{\micro\watt}$) for several metamaterial implementations of the annihilating slabs. (a)~Some of the input power is lost because of reflection on the metamaterial layer and because of scattering in leaky modes. Note also the amplified evanescent waves clearly visible in the gap between the metamaterial slabs. (b)~The black curve is the reference of bare silicon waveguides ($n = 3.476$). The blue curve with squares corresponds to a waveguide pair with left-handed claddings ($\mathrm{Re}[\epsilon] = \mathrm{Re}[\mu] = -1$, loss tangent = $10\%$). The red curves with circles, triangles, and diamonds show the resulting force for a poor man's implementation of the annihilating slabs in which the permittivity equals $-1$ and the loss tangent is $10\%$, $5\%$, and $1\%$, respectively.}
  \label{Fig4}
\end{figure}

In conclusion, our study shows that transformation optics allows for the design of materials which can significantly enhance the optical gradient force between two slab waveguides---the prototype system for optical gradient forces. When the space-annihilating waveguide cladding is implemented by a double-negative metamaterial, the force amplification is limited by dissipative loss. However, we can also use a single-negative metamaterial as a ``poor man's'' version of the annihilating medium. This approach results in much larger force enhancement factors while not impacting the mechanical properties of the waveguide system. We expect that further developments using hyperbolic media will be used to implement this concept at terahertz and optical frequencies. The principle outlined in this Letter may be generalized to other microphotonic force systems and, importantly, it may be combined with existing resonator-based enhancement methods \cite{VanThourhout-2010,Eichenfield-2007} to produce optical forces with unprecedented amplitude. This may lead to new ways of optical actuation in nano- and micrometer-scale devices.

Work at the Vrije Universiteit Brussel (computational studies) was supported by BelSPO (Grant IAP P7-35 photonics@be) and the Research Foundation--Flanders (FWO-Vlaanderen). Work at Ames Laboratory (theory) was supported by the U.S.\ Department of Energy, Office of Basic Energy Science, Division of Materials Sciences and Engineering (Ames Laboratory is operated for the U.S. Department of Energy by Iowa State University under Contract No. DE-AC02-07CH11358). V.\ G.\ acknowledges the Research Foundation--Flanders (FWO-Vlaanderen) for his Aspirant grant.


\end{document}